\documentclass[journal,comsoc]{IEEEtran}
\usepackage[T1]{fontenc}
\usepackage{cite}
\ifCLASSINFOpdf
\else
\fi
\usepackage{amsmath}
\usepackage{graphicx}
\usepackage{colortbl}
\usepackage{xcolor}
\usepackage[cmintegrals]{newtxmath}
\usepackage[utf8]{inputenc}
\begin{document}
\bibliographystyle{ieeetr}
\title{Codebook-Based Solutions for Reconfigurable Intelligent Surfaces and Their Open Challenges}%
\author{Jiancheng~An, Chao~Xu, Qingqing~Wu, Derrick~Wing~Kwan~Ng, Marco~Di~Renzo, Chau~Yuen, and~Lajos~Hanzo%
\thanks{J. An and C. Yuen are with the Engineering Product Development (EPD) Pillar, Singapore University of Technology and Design, Singapore 487372 (e-mail: jiancheng$\_$an@163.com; yuenchau@sutd.edu.sg). C. Xu and L. Hanzo are with the School of Electronics and Computer Science, University of Southampton, Southampton SO17 1BJ, U.K. (e-mail: cx1g08@soton.ac.uk; lh@ecs.soton.ac.uk). Q. Wu is with the Department of Electronic Engineering, Shanghai Jiao Tong University, Shanghai 200240, China (e-mail: wu.qq1010@gmail.com). D. W. K. Ng is with the School of Electrical Engineering and Telecommunications, University of New South Wales, Sydney, NSW 2052, Australia (e-mail: w.k.ng@unsw.edu.au). M. Di Renzo is with Universit\'e Paris-Saclay, CNRS, CentraleSup\'elec, Laboratoire des Signaux et Syst\`emes, 3 Rue Joliot-Curie, 91192 Gif-sur-Yvette, France (e-mail: marco.di-renzo@universite-paris-saclay.fr).}\vspace{-0.75cm}
}
\maketitle
\begin{abstract}
Reconfigurable intelligent surfaces (RIS) is a revolutionary technology to cost-effectively improve the performance of wireless networks. We first review the existing framework of channel estimation and passive beamforming (CE \& PBF) in RIS-assisted communication systems. To reduce the excessive pilot signaling overhead and implementation complexity of the CE \& PBF framework, we conceive a codebook-based framework to strike flexible tradeoffs between communication performance and signaling overhead. Moreover, we provide useful insights into the codebook design and learning mechanisms of the RIS reflection pattern. Finally, we analyze the scalability of the proposed framework by flexibly adapting the training overhead to the specified quality-of-service requirements and then elaborate on its appealing advantages over the existing CE \& PBF approaches. It is shown that our novel codebook-based framework can be beneficially applied to all RIS-assisted scenarios and avoids the curse of model dependency faced by its existing counterparts, thus constituting a competitive solution for practical RIS-assisted communication systems.
\end{abstract}
\IEEEpeerreviewmaketitle
\section{Introduction}\label{S1}
Over the past decades, channel fading has been considered one of the most unfavorable factors limiting the performance of wireless communication systems. In particular, the signal attenuation caused by large-scale fading -- including the distance-dependent path loss and shadow fading -- constrains both the service coverage range and communication quality, especially for the emerging millimeter-wave (mmWave) and TeraHertz (THz) bands. Although both the fine-grained massive multiple-input multiple-output (MIMO) beamforming and the classic amplify- or decode-and-forward relaying schemes constitute effective solutions, they generally result in extra hardware costs and exceedingly high power consumption. Furthermore, the small-scale fading leads to rapid fluctuation of the received signal power. In general, adaptive modulation and coding as well as resource allocation are adopted for mitigating the impact of small-scale fading. However, these classic designs are typically implemented at the transmitters or receivers \cite{Renzo_JSAC_2020_Smart}. As a consequence, the wireless channel is still treated as a random ``black box'' and handled proactively.

Recently, \emph{Reconfigurable Intelligent Surfaces (RIS)} have emerged as a promising technology for ameliorating wireless propagation environments in real-time owing to the quantum leap in research advances on programmable metasurfaces \cite{TWC_2022_Zhang_Dual}. Specifically, an RIS is a planar array comprising a large number of low-cost passive reflecting elements, each of which is capable of independently imposing a phase shift on the incident signals \cite{CM_2021_Pan_Reconfigurable}. From the perspective of large-scale fading, an RIS is capable of mitigating the path loss by partially reconstituting the electromagnetic waves diffused in space without requiring additional energy, which is a substantial benefit over conventional relaying and active beamforming. As for small-scale fading, an RIS is capable of advantageously manipulating the electromagnetic waves to establish constructive signal superposition, thus creating a more reliable communication channel. Alternatively, an RIS may be optimized to increase the effective rank of a communication link for acquiring much-needed spatial multiplexing gain \cite{TCOM_2022_An_Low}. In stark contrast to existing active transceiver-based link adaptive technologies, an RIS offers a new design degrees-of-freedom (DoF) for enhancing the performance attained, ushering in a new era of smart programmable wireless environments.
\begin{figure*}[!t]
	\centering
	\includegraphics[width=12cm]{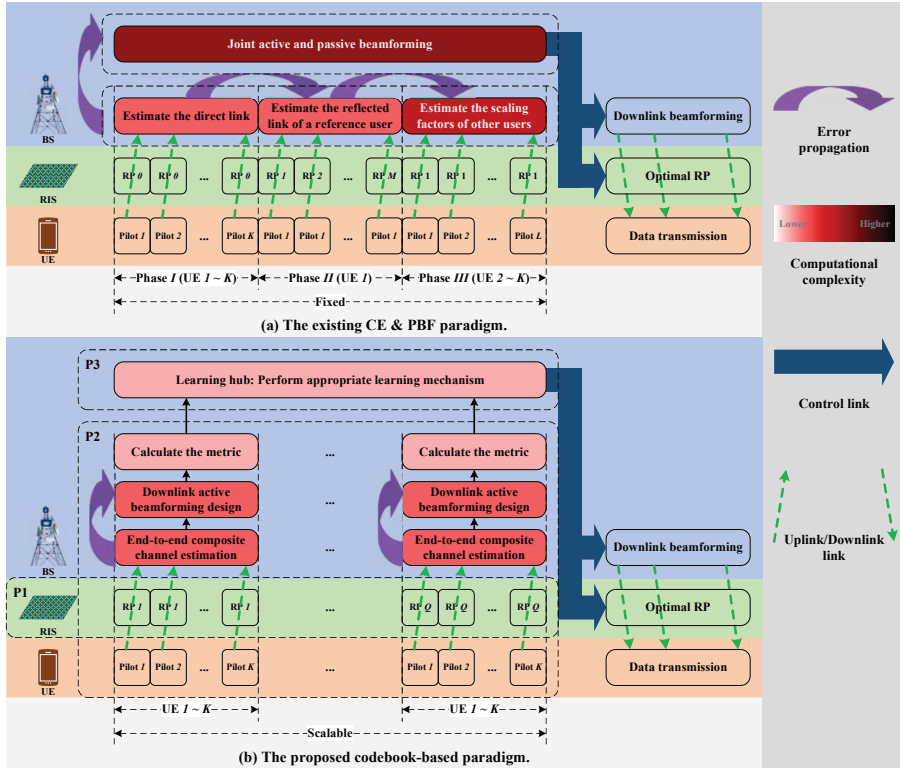}
	\caption{Protocol comparison between the existing CE \& PBF framework and the proposed codebook-based framework.}\vspace{-0.6cm}
	\label{fig1}
\end{figure*}

In order to unlock the full potential of RISs, channel estimation (CE) and passive beamforming (PBF) constitute a pair of crucial challenges. On the one hand, a large number of pilots are required for probing the numerous reflected links \cite{Jensen_ICASSP_2020_An}. Although advanced CE techniques have been conceived for reducing the pilot signaling overhead by exploiting the associated channel characteristics, the pilot overhead proportional to the number of RIS elements remains prohibitive \cite{Wang_TWC_2020_Channel}. On the other hand, most of the existing PBF designs pursue optimal performance by jointly optimizing the RIS and the active components in the network, which is nontrivial and requires bespoke solutions for different scenarios and applications \cite{Huang_TWC_2019_Reconfigurable, Wu_TWC_2019_Intelligent, Zhao_TWC_2021_intelligent}. Therefore, the scalability and complexity of the existing CE \& PBF frameworks constitute a stumbling block for the large-scale deployment of RISs, not to mention their heavy dependence on channel estimation \cite{WCL_2021_Gradoni_End}.

Against the above background, this article firstly offers a comprehensive review of the existing modularized CE \& PBF designs. Then, we discuss the critical issues of excessive pilot overhead and high implementation complexity. In order to improve both energy efficiency and scalability, we then propose a novel codebook-based framework for RIS-assisted communications, which efficiently addresses these challenges. Specifically, by first generating an appropriate RIS reflection pattern (RP) codebook offline, one only has to concentrate on the end-to-end composite channel estimation and the transmitter design, which can be directly inherited from current communication systems. Based on the labeled codebook, an appropriate learning-based approach is applied for determining the optimized RP for a given set of quality-of-service (QoS) requirements. Moreover, we design both the codebook as well as the relevant learning mechanisms and demonstrate the compelling benefits of the proposed codebook-based framework over the state-of-the-art.
\section{Existing CE \& PBF Frameworks}\label{S2}
In this section, we first survey the state-of-the-art CE \& PBF designs conceived for RIS-assisted communication systems and then pose several critical questions concerning the existing RIS-oriented frameworks.

\subsection{Channel Estimation (CE)}
\subsubsection{\underline{\textbf{Instantaneous CSI}}}
Channel state information (CSI) acquisition is extremely challenging for RIS-assisted systems because of the low-complexity passive nature of the RIS architecture. Specifically, due to the lack of sensing and advanced signal processing capability, it is unrealistic to separately estimate the base station (BS)-RIS and RIS-user equipment (UE) channels by harnessing traditional pilot-based techniques of active components. Instead, we opt for estimating the cascaded BS-RIS-UE channels \cite{Jensen_ICASSP_2020_An}. Specifically, upon repeatedly transmitting the same pilot and evaluating the received signal under different RPs, all the reflected channels can be obtained after collecting sufficient observations. Unfortunately, the number of pilots required is much higher than that of RIS elements, which is prohibitive for practical applications.

For the sake of reducing the pilot overhead, advanced CE techniques may be conceived by exploiting the specific channel characteristics. In \cite{Wang_TWC_2020_Channel}, a novel three-phase method was proposed for exploiting the commonality of the BS-RIS link among multiple UEs to reduce the pilot overhead. For the sake of illustration, Fig. \ref{fig1}(a) portrays this three-phase CE approach. In \emph{Phase I} consisting of $K$ slots, all $M$ RIS elements are switched off (i.e., RP $0$ in Fig. \ref{fig1}(a)) for estimating the direct channels associated with $K$ UEs. In \emph{Phase II} of Fig. \ref{fig1}(a), all the reflecting elements are switched on and merely a reference UE (assumed to be UE $1$) transmits pilot symbols to the BS. By leveraging the results of \emph{Phase I}, the reflected channels associated with the reference UE can be estimated within $M$ time slots. Finally, in \emph{Phase III} of Fig. \ref{fig1}(a), all other $\left( {K - 1} \right)$ UEs transmit their pilot symbols to the BS. Since the reflected channels of the $K$ UEs share the same BS-RIS link, the scaling factors between the different reflected channels are estimated instead of acquiring the full CSI, which requires at least $L =\max\left ( K-1,\left \lceil \left ( K-1 \right )M/N \right \rceil \right )$ slots with $N$ denoting the number of antennas at the BS. Interested readers may refer to \cite{Wang_TWC_2020_Channel} for the specific pilot design in \emph{Phase III}. In \cite{Zhao_TWC_2021_intelligent}, the different channel coherence time of the BS-RIS and RIS-UE links was exploited to further reduce the pilot overhead. Additionally, the exploitation of beamspace sparsity of mmWave communications and the spatial correlation between adjacent RIS elements can also mitigate the computational complexity as well as the pilot overhead at the cost of some performance erosion \cite{CM_2021_Pan_Reconfigurable}.

While pilot reduction has gradually become the prime objective of CE approaches conceived for RIS-assisted systems, their resultant performance erosion has been somewhat ignored \cite{Wang_TWC_2020_Channel}. Additionally, PBF has to be specifically tailored to accommodate the outputs of different CE approaches. In a nutshell, the existing modularized framework does not consider the intricate relationship between CE and PBF, it rather concentrates on the autonomous objective of each stage, which generally fails to optimize the holistic system performance at a given pilot overhead. For example, if there are only limited feasible RPs for $M$ reflecting elements due to practical hardware restrictions, it may become unnecessary to increase the pilot overhead proportionately with $M$. In contrast to the CE approaches that disregard the subsequent PBF requirements, it is a promising option to take into account the practical hardware constraints and ascertain what pivotal information is required for performing PBF.

\subsubsection{\underline{\textbf{Statistical CSI}}}
In order to reduce the number of pilots required by RIS-assisted systems, statistical CSI-based methods have also attracted substantial interests. The authors of \cite{Abrardo_TC_2021_Intelligent} maximized the sum rate of an RIS-assisted multi-user MIMO system by only exploiting the statistical location information of mobile users. Specifically, in the mmWave band, the channels of B5G/6G systems tend to largely rely on the line-of-sight (LoS) component, which allows aligning the RP with the pure LoS component \cite{TVT_2019_Han_Large}. As a benefit, the statistical CSI-based method significantly reduces the pilot overhead required for CSI acquisition; secondly, it also dispenses with frequent RIS reconfiguration as statistics vary slowly in time.

Although the statistical CSI-based method beneficially reduces the pilot overhead, it also has its disadvantages. Again, the statistical CSI-based method only performs well when strong LoS components are dominated. When the BS (UE) - RIS link is blocked, the statistical CSI-based method suffers from severe performance degradation. Furthermore, the statistical CSI-based scheme actually operates in a semi-static mode, thus reducing the DoF in optimizing the environment and failing to exploit the full potential of RIS. Therefore, striking a beneficial tradeoff between the reduced pilot overhead relying on statistical CSI and the excellent performance of instantaneous CSI requires future research.
\begin{table*}[!t]
	\centering
	\caption{A survey of existing CE \& PBF framework and codebook-based solutions for RIS-assisted communication systems.}
	\includegraphics[width=12cm]{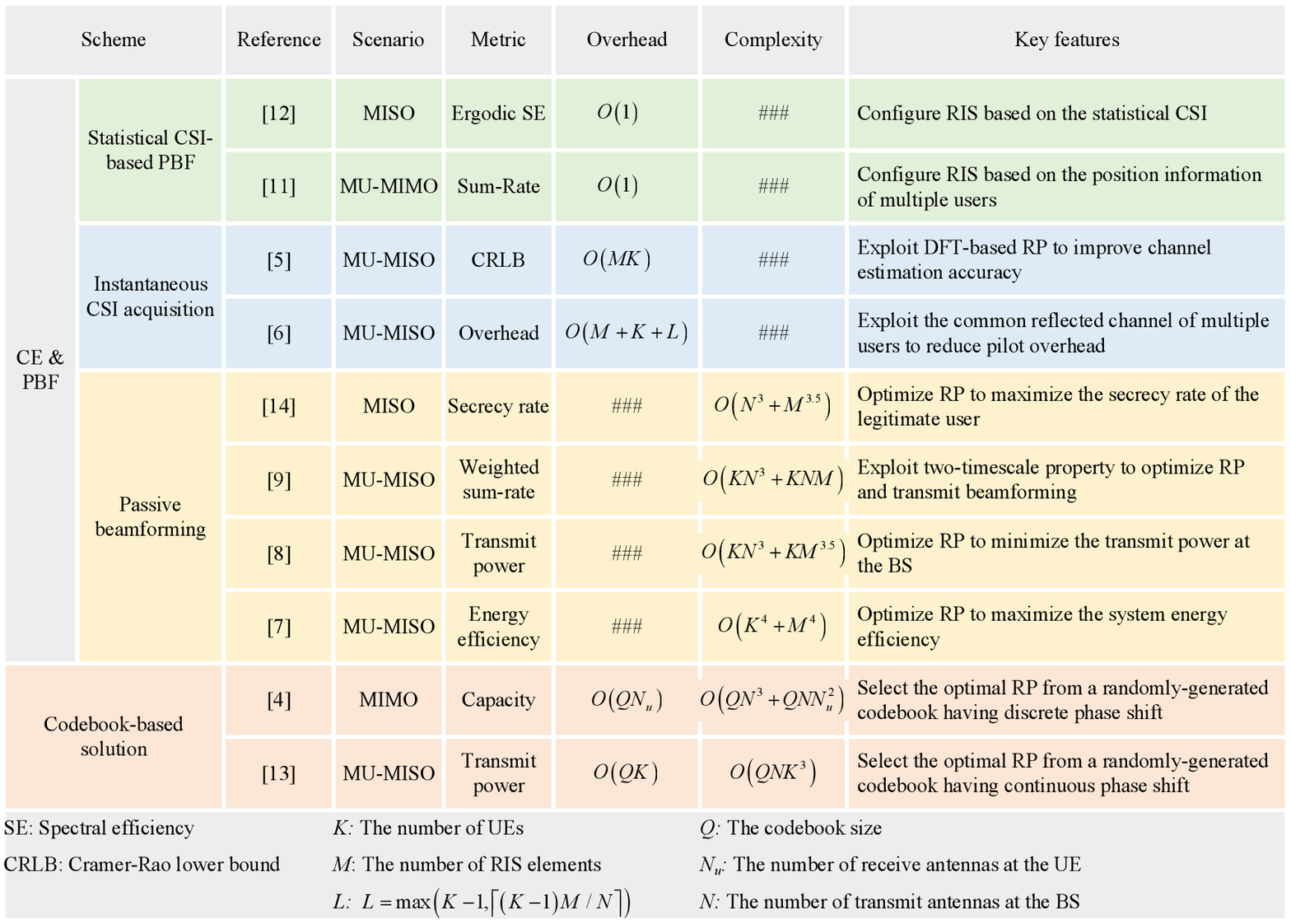}\vspace{-0.6cm}
	\label{tab0}
\end{table*}
\subsection{Passive Beamforming (PBF)}
PBF is another challenging task in RIS-assisted communications, which generally involves non-convex objective functions (OF) and constraints such as the discrete phase shifts \cite{CM_2021_Pan_Reconfigurable}. On the other hand, as shown in Fig. \ref{fig1}(a), the PBF at the RIS usually has to be designed in conjunction with the active components in the wireless network for gaining optimal performance \cite{Wu_TWC_2019_Intelligent}. A practical approach is to first relax the non-convex modulus constraints by applying, for example, successive convex approximation (SCA) or semi-definite relaxation (SDR), and then map the solution attained into its nearest legitimate solution \cite{Huang_TWC_2019_Reconfigurable}. Alternatively, the heuristic alternating optimization (AO) technique is capable of obtaining a sub-optimal solution with polynomial-time complexity by iteratively optimizing the active and passive components assuming that the other is fixed \cite{Wu_TWC_2019_Intelligent}. Indeed, the PBF design can be further simplified by optimizing the phase shifts of one RIS element each time, while fixing that of the rest until convergence is achieved. Upon leveraging the aforementioned optimization philosophy, RISs have already been shown to be of great potential in enhancing the channel quality, suppressing co-channel interference, as well as preventing eavesdropping \cite{TGCN_2022_An_Joint, Cui_WCL_2019_Secure}.

To elaborate, the quadratic power scaling law shows that the received power increases $6$ dB when doubling the number of reflecting elements in a far-field setting \cite{Wu_TWC_2019_Intelligent}. However, the conclusions of \cite{Wu_TWC_2019_Intelligent} are drawn based on the idealized assumption that all the reflected channels associated with different RIS elements are statistically independent, which requires adjacent RIS elements to be sufficiently far apart compared to the wavelength. In particular, the RIS's performance would be degraded by the channel's correlation and mutual coupling, which has to be quantitatively characterized by future research \cite{WCL_2021_Gradoni_End}. The calculus-based performance analysis is expected to be a powerful technique for characterizing the family of holographic RISs having infinitely densely deployed elements \cite{TWC_2022_Zhang_Dual}. In summary, the performance characterization of RISs depends on the underlying channel model, which requires further investigations and should also be verified by experimental measurements \cite{Renzo_JSAC_2020_Smart}.

\section{The Codebook-Based Solution}\label{S3}
\subsection{The Codebook-Based Protocol}\label{S3-1}
In this section, we develop a novel codebook-based framework for mitigating both the pilot overhead and the complexity. For the sake of illustration, Fig. \ref{fig1}(b) portrays the proposed codebook-based protocol. Instead of directly optimizing the RIS phase shifts, the codebook-based solution scans a limited set of RP candidates and learns the optimal RP based on the feedback from the controller. Thus, it becomes unnecessary to explicitly estimate the RIS-BS/UE channels. The detailed phases (\emph{\textbf{P}}) are summarized as follows:
\begin{itemize}
\item[\emph{\textbf{P1:}}] RIS appropriately extracts a codebook of RPs from the universal solution set taking into account the constraints imposed by the practical hardware structure, where the cardinality of the codebook is $Q$, as shown in Fig. \ref{fig1}(b).
\item[\emph{\textbf{P2:}}] For each RP in the codebook, all the UEs send uplink pilots and the BS estimates the end-to-end composite channel without explicitly considering the existence of the RIS. The number of pilots within each subframe (i.e., $K$ in Fig. \ref{fig1}(b)) depends on the total number of antennas at the UE side. Following this, the BS performs only the transmit beamformer optimization and evaluates the associated OF.
\item[\emph{\textbf{P3:}}] After repeating \emph{\textbf{P2}} for all RPs in the codebook, we now obtain $Q$ labeled training samples, based on which the BS executes the appropriate learning algorithm. Finally, for a given input of QoS target, the learning system would return an optimal RP for downlink data transmission. Moreover, new observations can be employed during the data transmission phase to further improve the learning accuracy.
\end{itemize}

Before proceeding further, we first contrast the proposed codebook-based framework with existing CE \& PBF designs. First of all, the codebook-based solution estimates the end-to-end composite channel at each time slot, which is exactly what both the transmitter and receiver need and thus avoids the potential pilot wastage of traditional approaches. The rest of the design problem is to adjust the size of the codebook according to the specific scenarios for striking beneficial tradeoffs between the pilot overhead and the ultimate communication performance. As for the PBF design, the codebook is generated directly based on the legitimate RP set, which implicitly takes all the adverse practical factors into account during its learning process. By contrast, existing designs aim for the optimal solution under various idealized assumptions, e.g., the perfect CSI and flawless element tuning, which would result in severe mismatch errors, when considering the hardware imperfections. Although the CE \& PBF framework can also take any adverse constraints into account at the beginning of the problem formulation, it has to be regularly updated to cater for various scenarios. For the sake of illustration, we boldly and explicitly contrast the codebook-based solution to its existing CE \& PBF counterparts in Table \ref{tab0}.

\begin{figure*}[!t]
	\centering
	\includegraphics[width=11cm]{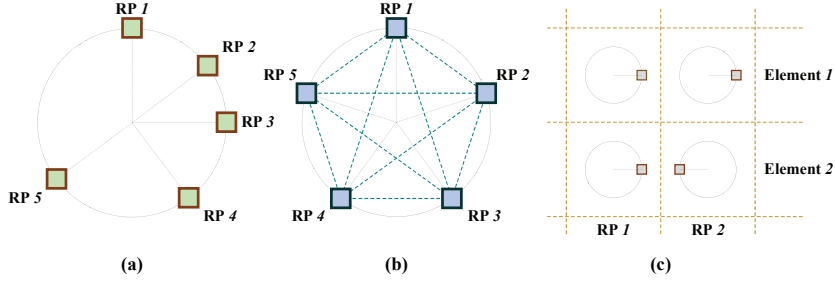}
	\caption{The schematic of three RP codebooks: (a) Random ($M =1,\ Q = 5$); (b) SDM ($M = 1,\ Q = 5$); (c) Orthogonal ($M = 2,\ Q = 2$).}
	\label{fig2}
\end{figure*}
\begin{figure*}[!t]
	\centering
	\includegraphics[width=11cm]{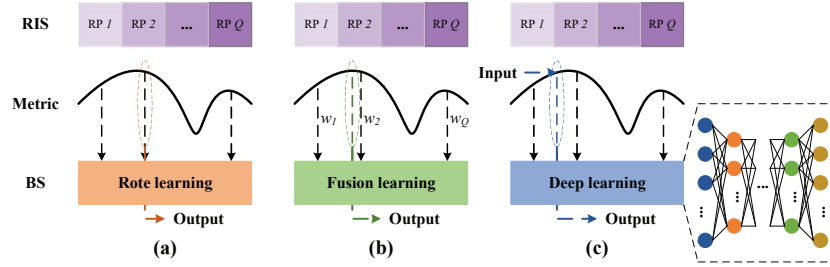}
	\caption{The schematic of three learning mechanisms: (a) Rote learning; (b) Fusion learning; (c) Deep learning.}\vspace{-0.6cm}
	\label{fig3}
\end{figure*}
\subsection{Benefits of the Codebook-Based Framework}\label{S4-2}
Next, we elaborate on the advantages of the proposed codebook-based framework over the existing CE \& PBF framework from the following six aspects.

\subsubsection{\underline{\textbf{Backward Compatibility}}}
The proposed codebook-based framework can be directly applied to the conventional communication architecture for optimizing the RIS benefits \cite{TCOM_2022_An_Low}. The only additional configuration requirement is to appropriately increase the training overhead. During each training period, the BS repeats the end-to-end channel estimation and transmitter training in the same way as the existing framework. By contrast, the cascaded CE \& PBF framework tends to replace the entire suite of existing protocols and thus has to be reformulated for different scenarios and setups \cite{Huang_TWC_2019_Reconfigurable}.

\subsubsection{\underline{\textbf{Scalable Pilot Overhead}}}
As shown in Fig. \ref{fig1}(b), the proposed scheme is capable of adjusting the training overhead according to the channel's coherence time so that the pilot overhead can be flexibly adapted to the specific scenario. This is in sharp contrast to the existing CE schemes \cite{Jensen_ICASSP_2020_An, Wang_TWC_2020_Channel}, which require an excessive pilot signaling overhead for probing a large number of reflected channels. Although the progressive CE schemes of the conventional framework attempt to minimize the required pilot signaling overhead, they remain dependent on the number of reflecting elements, which essentially limits the practical application of RISs \cite{TCOM_2021_Najafi_Physics}.

\subsubsection{\underline{\textbf{Reduced Computational Complexity}}}
Compared to the computationally inefficient joint active and passive beamforming design, the codebook-based solution substantially reduces the system complexity by decoupling the optimization of PBF and that of the active elements, as shown in \emph{\textbf{P2}} of Fig. \ref{fig1}(b). Moreover, when considering practical hardware imperfections, the reduced complexity may not necessarily lead to an inferior performance compared to the existing schemes relying on unrealistic idealized assumptions, especially in the case of large-scale deployments.

\subsubsection{\underline{\textbf{Reduced Error Propagation}}}
In the existing CE \& PBF frameworks, the successive estimation of different channels, the separate design of CE and PBF, as well as the mismatch between the practical RIS model and its relaxed counterpart will lead to error propagation (see Fig. \ref{fig1}(a)), thereby degrading the expected performance \cite{Wang_TWC_2020_Channel}. By contrast, in the codebook-based framework, error propagation only occurs at the active transmit beamformer design due to the imperfect CSI of the end-to-end composite channel, and its negative impact can be further mitigated by applying joint CE and transmit beamformer optimization. Hence, the proposed codebook-based framework substantially mitigates the error propagation compared to the existing CE \& PBF design.

\subsubsection{\underline{\textbf{Reduced Control Signaling}}}
The proposed codebook-based solution effectively reduces the capacity requirement of the control link. Assuming $B$ legitimate phase shifts for each RIS element, the existing methods require $M{\log _2}B$ signaling bits for configuring $M$ RIS elements. By contrast, the control signaling overhead of our codebook-based solution is only ${\log _2}Q$ bits. Therefore, the proposed scheme substantially reduces the backhaul overhead and delay.

\subsubsection{\underline{\textbf{Stronger Robustness}}}
The proposed codebook-based scheme estimates the composite end-to-end channel without considering the individual channels reflected via the RIS. Specifically, the channel correlation, mutual coupling, component aging, and other adverse factors would result in severe performance degradation for the existing CE \& PBF solution, which has not been fully characterized \cite{TGCN_2022_An_Joint}. By contrast, all the aforementioned hardware imperfections have been implicitly taken into consideration during the on-site learning process of our codebook-based scheme, while the existing solutions would have to be redesigned or regularly updated for different scenarios. Therefore, the proposed scheme is more robust to hardware imperfections.

In summary, the existing CE \& PBF framework can be characterized as \emph{``sensing first and then configuring''}, whereas the proposed codebook-based solution is based on the \emph{``learning from testing''} philosophy.

\section{Research Opportunities and Challenges}
Naturally, the specific generation of the RP codebook and the learning algorithm determine the performance attained, hence they are still under investigation, as discussed below.
\subsection{Codebook Generation}
Here we highlight three methods of generating the RP codebook, i.e., \textbf{\emph{P1}} in Fig. \ref{fig1}(b). 
\subsubsection{\underline{\textbf{Random Codebook}}}
First of all, let us consider the codebook having $Q$ identical RPs. Thus, we obtain $Q$ observations of the same channel, which only attains a mean square error (MSE) performance gain of $10\log_{10}Q$ dB for CE. However, by beneficially designing a codebook containing different RPs, one would obtain extra diversity gain. To this end, the most straightforward method is to randomly generate the RP codebook, where each element in the codebook is randomly selected from the universal set of RPs \cite{TCOM_2022_An_Low}. For the sake of illustration, Fig. \ref{fig2}(a) shows a randomly generated codebook considering $M=1$ and $Q=5$.

\subsubsection{\underline{\textbf{Sum Distance Maximized (SDM) Codebook}}}
It is evident that the randomly generated codebook of RPs does not exploit the limited training overhead effectively. To circumvent this issue, the authors of \cite{TGCN_2022_An_Joint} conceived a heuristic method of generating $Q$ different RPs by maximizing the sum of Euclidean distances between all pairs of RPs, as shown in Fig. \ref{fig2}(b). The SDM codebook has been shown to achieve beneficial received power gains. Nevertheless, since the SDM codebook cannot examine all the legitimate RPs, it only provides moderate gain compared to the random codebook \cite{TGCN_2022_An_Joint}. Therefore, how to generate an appropriate codebook conducive to subsequent learning for adapting to a variety of application scenarios and optimization objectives remains an open question.
\subsubsection{\underline{\textbf{Orthogonal Codebook}}}
In conventional beam control and tracking regimes, employing orthogonal beam codebooks (e.g., the DFT-based steering vector) has become a popular option. Hence in RIS-assisted communication systems, DFT-based RP design has also been employed for improving the estimation accuracy \cite{Wang_TWC_2020_Channel}, but a broader family of orthogonal codebooks as well as the resultant performance analysis deserves further exploration. Specifically, an orthogonal codebook may be used for searching for the LoS component at a low overhead \cite{TCOM_2022_An_Low}. In this context Fig. \ref{fig2}(c) shows an orthogonal codebook considering $M=2$ and $Q=2$.

Note that in addition to the aforementioned environment-agnostic codebook designs, one could exploit the statistical CSI to design a CSI-aware codebook for further improving the performance at the expense of increasing the backhaul capacity. Furthermore, beam training could also be improved by dividing the potential service zone into several beams in a hierarchical manner \cite{TWC_2022_Zhang_Dual}.
\subsection{Learning Mechanism}
Then, a set of three learning mechanisms is introduced to perform PBF based on $Q$ observations of the end-to-end composite channel~\cite{TCOM_2022_An_Low, TGCN_2022_An_Joint}, i.e., \textbf{\emph{P3}} in Fig.~\ref{fig1}(b).
\subsubsection{\underline{\textbf{Rote Learning}}}
Fig. \ref{fig3}(a) portrays the most straightforward learning method, namely rote learning \cite{TGCN_2022_An_Joint}, which is the most classic codebook-based scheme relying on memorizing multiple observations. Explicitly, the BS chooses the specific RP that performs the best in terms of the OF from the codebook, as usual. Thus, the control signaling overhead of $\log_{2}Q$ bits is required to configure the RIS.
\subsubsection{\underline{\textbf{Fusion Learning}}}
Another feasible technique is fusion learning, which synthesizes more reliable statistical inferences from multiple observations. As shown in Fig. \ref{fig3}(b), by weighting and superimposing different RPs in the pre-designed codebook using a set of weighting coefficients $w_1,\, w_2,\, \cdots, \, w_{Q}$, one may attain a better performance by making each weighting coefficient proportional to the corresponding OF value. Naturally, the resultant performance is heavily dependent on the OF. Specifically, fusion learning will outperform rote learning for a convex optimization problem at the expense of increasing the backhaul capacity from $\log_{2}Q$ bits to $Q$ bits \cite{TCOM_2022_An_Low}. However, it is non-trivial to obtain a unified fusion learning framework suitable for all OFs. As such, the combination with some convex relaxation techniques, e.g., SCA, may be beneficially harnessed, but this requires further investigations.

\subsubsection{\underline{\textbf{Machine Learning}}}
Machine learning (ML) techniques, especially deep learning, can be efficiently applied in our codebook-based framework instead of the aforementioned model-based methods. As shown in Fig. \ref{fig3}(c), the codebook is first labeled upon calculating the OF value for each RP. Following this, the performance metrics and corresponding reflection coefficients are used to train a deep neural network (DNN) as input and output, respectively. Then the desired QoS requirement is used as input and a well-trained DNN is utilized to return an appropriate RP to assist the data transmission. A series of new data can be employed for further training the DNN to improve its performance during the testing phase \cite{Renzo_JSAC_2020_Smart}. In contrast to the above two model-based learning mechanisms, which solve the associated maximization/minimization problems heuristically, the data-driven DNN introduced here is supervised with the aid of labeled samples, which aims for returning an optimized RP for satisfying the user's expected QoS target. Although advanced deep reinforcement learning (DRL) can be employed for updating the RP in real-time, more advanced unsupervised ML techniques capable of completely dispensing with labeled data for solving these challenging optimization problems require further exploration.

\section{Performance Evaluation}\label{S4}
\subsection{Fundamental Tradeoff: Performance vs. Pilot Overhead}
As expected, there is a fundamental performance \emph{vs.} pilot overhead tradeoff in RIS-assisted systems. Specifically, the RIS performance relies on the CSI accuracy and hence on the pilot overhead. In Table \ref{tab1}, we first illustrate the stylized performance versus overhead tradeoff for providing insights into the scalability of the codebook-based framework, where the rote learning mechanism is adopted. Explicitly, the performance characterized in Table \ref{tab1} relies on theoretical evaluation and it is validated in terms of the received power scaling law (i.e., $P_{r}$) \cite{TGCN_2022_An_Joint, TCOM_2022_An_Low}. Observe from Table \ref{tab1} that the optimal RIS performance is characterized in terms of quadratic gain \emph{vs.} the number of RIS elements \cite{Wu_TWC_2019_Intelligent}, but this may only be attained at a proportionately increased training overhead. By contrast, the codebook-based solution is capable of striking a compelling performance vs. overhead trade-off by flexibly adjusting $Q$. When considering the maximum codebook size of $Q_{\max} = B^{M}$ with $B$ denoting the number of phase shifts, we have ${P_r} \propto O\left( {{M^2}\log B} \right)$ for the random codebook, which implies that the quadratic power scaling law could be achieved by exhaustively searching through the entire codebook space. Furthermore, the improved SDM codebook is capable of achieving a quadratic gain, but it suffers from some performance erosion determined by the codebook size. More advanced codebooks and learning mechanisms could achieve improved performance despite a limited pilot signaling overhead, but the associated performance evaluation is still an open research area.
\begin{table}[!t]
	\centering
	\caption{Overhead and performance comparison between codebook-based framework and its existing counterparts.}
	\includegraphics[width=6.5cm]{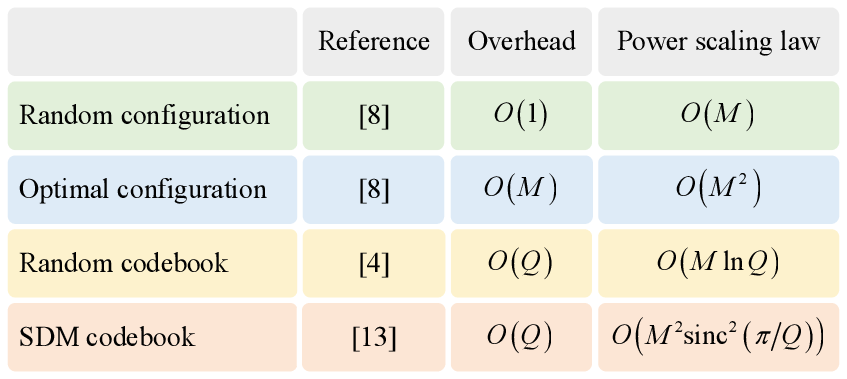}\vspace{-0.6cm}
	\label{tab1}
\end{table}

\subsection{Case Study}
\begin{figure*}[!t]
	\centering
	\includegraphics[width=13cm]{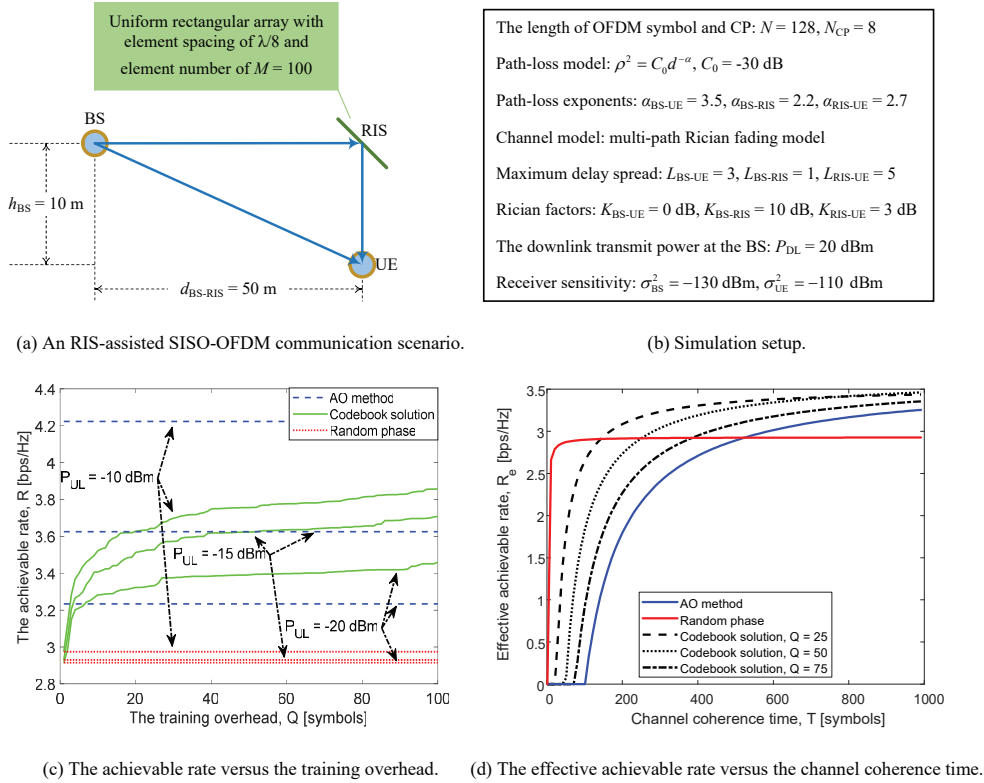}
	\caption{Case study by considering an RIS-assisted SISO-OFDM system.}\vspace{-0.6cm}
	\label{fig5}
\end{figure*}

In this section, we provide some numerical results to show the benefits of the codebook-based solution. Given the strict page limit, we only employ the random codebook as well as the rote learning mechanism. In contrast to \cite{TGCN_2022_An_Joint}, we consider an RIS-assisted single-input single-output orthogonal frequency division multiplexing (SISO-OFDM) communication system, as shown in Fig. \ref{fig5}(a). All the parameters in our simulations are listed in Fig. \ref{fig5}(b).

First, we study the rate performance of the codebook-based solution under imperfect CSI in Fig. \ref{fig5}(c), where the AO method \cite{Wu_TWC_2019_Intelligent} and the random phase shift are adopted as benchmark schemes. The DFT-based RP is employed for improving the uplink channel estimation accuracy. Observe from Fig. \ref{fig5}(c) that the codebook-based scheme degrades and essentially it becomes the random phase shift based regime when $Q=1$. Furthermore, compared to the AO method, the codebook-based scheme suffers from some performance penalty in the case of a high pilot power, but it outperforms the AO method in the prevalent practical scenario of having low pilot power at the UE side and an adequate number of training slots (e.g., $P_{\text{UL}} = -20$ dBm and $Q > 10$). In a nutshell, the estimation errors of the direct channels and all reflected channels severely degrade the existing CE \& PBF performance. By contrast, the codebook-based scheme is shown to be more robust against imperfect CSI.

Then Fig. \ref{fig5}(d) shows the effective achievable rate versus the channel's coherence time with $P_{\text{UL}} = -15$ dBm. Specifically, the effective achievable rate is defined by $R_{\text{e}} = (1-\tau/T)R$, where $\tau$ and $T$ represent the training overhead and the channel coherence time, respectively. As seen from Fig. \ref{fig5}(d), when considering rapidly time-varying channels having short coherence time (e.g., $T<500$), the AO method performs quite poorly in terms of its effective achievable rate, which is a consequence of its excessive overhead required for probing all the reflected channels. By contrast, the random phase shift having the lowest overhead results in the highest effective achievable rate. Fortunately, the scalable codebook-based scheme can dynamically adapt the training overhead $Q$, which may be optimized for maximizing the effective achievable rate vs the channel coherence time.

\section{Conclusions}\label{S5}
A promising codebook-based framework was proposed for RIS-assisted systems to strike flexible performance and pilot overhead tradeoffs. Specifically, our codebook-based solution is demonstrated to be more suitable for ultra-dense wireless networks with large-scale RIS deployment. Moreover, we discussed the fundamental issues of the codebook-based framework, including the generation of the codebook as well as properly designed learning methods, and elaborated on its benefits over existing CE \& PBF solutions. Since the codebook-based framework remains largely unexplored, this article will provide useful guidance for unleashing the full potential of RIS in practical applications of future wireless networks.
\section*{Acknowledgment}
This research is supported by the Ministry of Education, Singapore, under its MOE Tier 2 (Award number MOE-T2EP50220-0019). Any opinions, findings and conclusions or recommendations expressed in this material are those of the author(s) and do not reflect the views of the Ministry of Education, Singapore. D. W. K. Ng is supported by the Australian Research Council's Discovery Project (DP210102169). The work of M. Di Renzo was supported in part by the European Commission through the H2020 ARIADNE project under grant agreement number 871464 and through the H2020 RISE-6G project under grant agreement number 101017011. L. Hanzo would like to acknowledge the financial support of the Engineering and Physical Sciences Research Council projects EP/W016605/1 and EP/P003990/1 (COALESCE) as well as of the European Research Council's Advanced Fellow Grant QuantCom (Grant No. 789028).
\bibliography{ref}

\begin{thebibliography}{10}

\bibitem{Renzo_JSAC_2020_Smart}
M.~Di~Renzo, \emph{et al.}, ``Smart radio environments empowered by
  reconfigurable intelligent surfaces: How it works, state of research, and the
  road ahead,'' {\em IEEE J. Sel. Areas Commun.}, vol.~38, pp.~2450--2525, Nov.
  2020.

\bibitem{TWC_2022_Zhang_Dual}
Y.~Zhang, \emph{et al.}, ``Dual codebook design for intelligent omni-surface
  aided communications,'' {\em IEEE Trans. Wireless Commun.}, pp.~1--14, 2022,
  Early Access.

\bibitem{CM_2021_Pan_Reconfigurable}
C.~Pan, \emph{et al.}, ``Reconfigurable intelligent surfaces for {6G} systems:
  Principles, applications, and research directions,'' {\em IEEE Commun. Mag.},
  vol.~59, pp.~14--20, Jun. 2021.

\bibitem{TCOM_2022_An_Low}
J.~An, \emph{et al.}, ``Low-complexity channel estimation and passive
  beamforming for {RIS}-assisted {MIMO} systems relying on discrete phase
  shifts,'' {\em IEEE Trans. Commun.}, vol.~70, pp.~1245--1260, Feb. 2022.

\bibitem{Jensen_ICASSP_2020_An}
T.~L. Jensen and E.~De~Carvalho, ``An optimal channel estimation scheme for
  intelligent reflecting surfaces based on a minimum variance unbiased
  estimator,'' in {\em Proc. IEEE ICASSP}, pp.~5000--5004, May 2020.

\bibitem{Wang_TWC_2020_Channel}
Z.~Wang, L.~Liu, and S.~Cui, ``Channel estimation for intelligent reflecting
  surface assisted multiuser communications: Framework, algorithms, and
  analysis,'' {\em IEEE Trans. Wireless Commun.}, vol.~19, pp.~6607--6620, Oct.
  2020.

\bibitem{Huang_TWC_2019_Reconfigurable}
C.~Huang, \emph{et al.}, ``Reconfigurable intelligent surfaces for energy
  efficiency in wireless communication,'' {\em IEEE Trans. Wireless Commun.},
  vol.~18, pp.~4157--4170, Aug. 2019.

\bibitem{Wu_TWC_2019_Intelligent}
Q.~Wu and R.~Zhang, ``Intelligent reflecting surface enhanced wireless network
  via joint active and passive beamforming,'' {\em IEEE Trans. Wireless
  Commun.}, vol.~18, pp.~5394--5409, Nov. 2019.

\bibitem{Zhao_TWC_2021_intelligent}
M.-M. Zhao, \emph{et al.}, ``Intelligent reflecting surface enhanced wireless
  networks: Two-timescale beamforming optimization,'' {\em IEEE Trans. Wireless
  Commun.}, vol.~20, pp.~2--17, Jan. 2021.

\bibitem{WCL_2021_Gradoni_End}
G.~Gradoni and M.~Di~Renzo, ``End-to-end mutual coupling aware communication
  model for reconfigurable intelligent surfaces: An electromagnetic-compliant
  approach based on mutual impedances,'' {\em IEEE Wireless Commun. Lett.},
  vol.~10, pp.~938--942, May 2021.

\bibitem{Abrardo_TC_2021_Intelligent}
A.~Abrardo, D.~Dardari, and M.~Di~Renzo, ``Intelligent reflecting surfaces:
  Sum-rate optimization based on statistical position information,'' {\em IEEE
  Trans. Commun.}, vol.~69, pp.~7121--7136, Oct. 2021.

\bibitem{TVT_2019_Han_Large}
Y.~Han, \emph{et al.}, ``Large intelligent surface-assisted wireless
  communication exploiting statistical {CSI},'' {\em IEEE Trans. Veh.
  Technol.}, vol.~68, pp.~8238--8242, Aug. 2019.

\bibitem{TGCN_2022_An_Joint}
J.~An, \emph{et al.}, ``Joint training of the superimposed direct and reflected
  links in reconfigurable intelligent surface assisted multiuser
  communications,'' {\em IEEE Trans. Green Commun. Netw.}, vol.~6,
  pp.~739--754, Jun. 2022.

\bibitem{Cui_WCL_2019_Secure}
M.~Cui, G.~Zhang, and R.~Zhang, ``Secure wireless communication via intelligent
  reflecting surface,'' {\em IEEE Wireless Commun. Lett.}, vol.~8,
  pp.~1410--1414, Oct. 2019.

\bibitem{TCOM_2021_Najafi_Physics}
M.~Najafi, \emph{et al.}, ``Physics-based modeling and scalable optimization of
  large intelligent reflecting surfaces,'' {\em IEEE Trans. Commun.}, vol.~69,
  pp.~2673--2691, Apr. 2021.

\end{thebibliography}
\begin{IEEEbiographynophoto}{Jiancheng An}[M'22]
received the B.S. degree in Electronics and Information Engineering and the Ph.D. degree in Information and Communication Engineering from the University of Electronic Science and Technology of China (UESTC), Chengdu, China, in 2016 and 2021, respectively. From 2019 to 2020, he was a Visiting Scholar with the Next-Generation Wireless Group, University of Southampton, U.K. He is currently a research fellow with the Engineering Product Development (EPD) Pillar, Singapore University of Technology and Design (SUTD). His research interests include reconfigurable intelligent surfaces (RIS), and integrated sensing and communications (ISAC).
\end{IEEEbiographynophoto}
\begin{IEEEbiographynophoto}{Chao Xu}[SM’19]
received a B.Eng. degree from Beijing University of Posts and Telecommunications, China, and a BSc(Eng) with First Class Honours from Queen Mary, University of London, UK, through a Sino-UK joint degree program in 2008, both in Telecommunications. He obtained a MSc degree with distinction in Radio Frequency Communication Systems and a Ph.D. degree in Wireless Communications from the University of Southampton, UK, in 2009 and 2015, respectively. He is currently a senior research fellow working at Next Generation Wireless Research Group, University of Southampton, UK. His research interests include index modulation, reconfigurable intelligent surfaces, noncoherent detection and turbo detection. He was awarded the Best M.Sc. Student in Broadband and Mobile Communication Networks by the IEEE Communications Society (United Kingdom and Republic of Ireland Chapter) in 2009. He also received 2012 Chinese Government Award for Outstanding Self-Financed Student Abroad and 2017 Dean’s Award, Faculty of Physical Sciences and Engineering, the University of Southampton.
\end{IEEEbiographynophoto}
\begin{IEEEbiographynophoto}{Qingqing Wu}[SM’21]
was listed as the Clarivate ESI Highly Cited Researcher in 2022 and 2021, the Most Influential Scholar Award in AI-2000 by Aminer in 2021 and World’s Top 2\% Scientist by Stanford University in 2020 and 2021. He was the recipient of the IEEE ComSoc Asia Pacific Best Young Researcher Award and Outstanding Paper Award in 2022, the IEEE ComSoc Young Author Best Paper Award in 2021.
\end{IEEEbiographynophoto}
\begin{IEEEbiographynophoto}{Derrick Wing Kwan Ng}[F'21]
received his Ph.D. degree from the University of British Columbia in 2012. He is now working as an Associate Professor at the University of New South Wales, Sydney, Australia. His research interests include convex and non-convex optimization, physical layer security, wireless information and power transfer, and green (energy-efficient) wireless communications.
\end{IEEEbiographynophoto}
\begin{IEEEbiographynophoto}{Marco Di Renzo}[F'20]
is a CNRS Research Director (Professor) and the Head of the Intelligent Physical Communications in the Laboratory of Signals and Systems at Paris-Saclay University - CNRS and CentraleSupelec. He serves as the Coordinator of the Communications and Networks Area of the Laboratory of Excellence DigiCosme, and as a Member of the Admission and Evaluation Committee of the Ph.D. School of Paris-Saclay University. He is a Fulbright Fellow at City University of New York, USA; an Fellow of IEEE, IET, AAIA, Vebleo; an Ordinary Member of EASA and the Academia Europaea; and a Highly Cited Researcher. He received the 2022 Michel Monpetit Prize from the French Academy of Sciences. He serves as the Editor-in-Chief of IEEE Communications Letters.
\end{IEEEbiographynophoto}
\begin{IEEEbiographynophoto}{Chau Yuen}[F’21]
received the B.Eng. and Ph.D. degrees from Nanyang Technological University (NTU), Singapore, in 2000 and 2004, respectively. He was a Post-Doctoral Fellow with Lucent Technologies Bell Labs, Murray Hill, in 2005. From 2006 to 2010, he was with the Institute for Infocomm Research (I2R), Singapore. Since 2010, he has been with the Singapore University of Technology and Design. Dr. Yuen was a recipient of the Lee Kuan Yew Gold Medal, the Institution of Electrical Engineers Book Prize, the Institute of Engineering of Singapore Gold Medal, the Merck Sharp and Dohme Gold Medal, and twice a recipient of the Hewlett Packard Prize. He received the IEEE Asia Pacific Outstanding Young Researcher Award in 2012 and IEEE VTS Singapore Chapter Outstanding Service Award on 2019. Currently, he serves as an Editor for the IEEE TRANSACTIONS ON VEHICULAR TECHNOLOGY, IEEE System Journal, and IEEE Transactions on Network Science and Engineering. He served as the guest editor for several special issues, including IEEE JOURNAL ON SELECTED AREAS IN COMMUNICATIONS, IEEE WIRELESS COMMUNICATIONS MAGAZINE, IEEE TRANSACTIONS ON COGNITIVE COMMUNICATIONS AND NETWORKING. He is a Distinguished Lecturer of IEEE Vehicular Technology Society.
\end{IEEEbiographynophoto}
\begin{IEEEbiographynophoto}{Lajos Hanzo}[F'04]
(http://www-mobile.ecs.soton.ac.uk, https://en.wikipedia.org/wiki/Lajos\_Hanzo) received his Master degree and Doctorate in 1976 and 1983, respectively from the Technical University (TU) of Budapest. He was also awarded the Doctor of Sciences (DSc) degree by the University of Southampton (2004) and Honorary Doctorates by the TU of Budapest (2009) and by the University of Edinburgh (2015). He is a Foreign Member of the Hungarian Academy of Sciences and a former Editor-in-Chief of the IEEE Press. He has served several terms as Governor of both IEEE ComSoc and of VTS. He has published widely at IEEE Xplore, 19 Wiley-IEEE Press books and has helped the fast-track career of 123 PhD students. Over 40 of them are Professors at various stages of their careers in academia and many of them are leading scientists in the wireless industry. He is also a Fellow of the Royal Academy of Engineering (FREng), of the IET and of EURASIP. He holds the Eric Sumner Field Award.
\end{IEEEbiographynophoto}
\end{document}